\documentclass[aps,prl,showpacs,twocolumn,groupedaddress,amssymb]{revtex4}

\usepackage{graphicx}
\usepackage{dcolumn}
\usepackage{bm}

\begin{document}

\title{THz light source Based on the two-stream instability and Backward Raman Scattering}

\author{S. Son}
\affiliation{169 Snowden Lane, Princeton, NJ, 08540}

\begin{abstract}
A new scheme of the THz light source is proposed. The excitation of the Langmuir wave in a moderately relativistic electron beam via the two-stream instability and the subsequent interaction of the excited Langmuir waves with the visible light-laser results in THz light via the Raman scattering. The cost of the current scheme could be cheaper than the conventional technologies by using the electron beam from cathodes and low-intensity infra-red laser.
\end{abstract}
\maketitle
 
\textit{An intense THz light source} could create new industries in the biomedical image,  the molecular spectroscopy, the tele-communication and many others~\cite{diagnostic, siegel,siegel2,  siegel3, booske,radar, security}.
If commercial viable intense THz light sources were available,
its scientific and practical implact would be immense. 
While  many THz light sources have been invented~\cite{Tilborg,Zheng,Reimann, gyrotron, tgyro, gyrotron2, gyrotron3,magnetron, qlaser, qlaser3, freelaser, freelaser2, colson, songamma,Gallardo,sonttera},
  a few technological difficulties persist in achieving a THz light intense enough for the purpose of mentioned  applications.
The current inability to produce high intense THz light comparable to the theoretical limit is referred to as the ``THz gap''~\cite{siegel, siegel2, booske}, which is  the major hurdle in commercializing THz light sources.
Another serious difficulty in realizing the plausible applications of intense THz light is  the high operating and construction cost;  Becuase current THz light sources often need expensive strong magnetic field or accelerators, and  often need to be operatured in extremly low-temperautre, the operating (construction) cost is very high. The conversion efficiency from the input energy into the THz light,  currently as low as 0.01 percents, renders the operating cost even higher.  Therefore, any progress in enhancing the intensity (the power) or reducing the cost (size) of a  THz light source is critical. 

\begin{figure}
\scalebox{0.3}{
\includegraphics[width=1.7\columnwidth, angle=270]{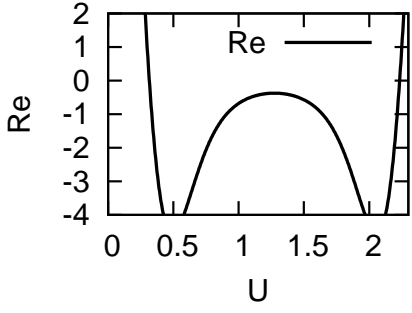}}
\caption{\label{fig1}
The real part of the dielectric function $\epsilon $ as a function of the frequency for the classical plasmas. The x-axis is $\mathrm{U} = \omega / \omega_{pe} $ and the y-axis is $\mathrm{Re} = \mathrm{Re}[\epsilon(\omega)] $. In this example, $n_0 =   10^{15} / \mathrm{cc} $, $\gamma_0 =1.5$, $T_e = 50 \  \mathrm{eV} $, $v_0 /c = 0.08 $, $k \lambda_{de} = 0.35 $ so that $k v_0 /\omega_{pe} \cong = 2.83 $. The local maxima of the real part at $ \omega = 1.5  \ \omega_{pe} $ is less than 0, which  is the threshold condition of the two-stream instability. 
}
\end{figure}

In this paper, the author proposes a new THz light source, by exciting the Langmuir waves via the two-stream instability inside an electron beam and emitting THz light via the backward Raman scattering (BRS) between the excited Langmuir waves and an infra-red laser.   The infra-red laser propagates with the electron beam in the same direction and the THz light is emitted in the opposite direction to the electron beam as shown in Fig.~(\ref{fig1}).
 The current scheme removes  the most important difficulty in the BRS-based THz light source; generating a strong seed THz pulse. The seed THz light  is very hard to generate but pre-requisite for the THz light generation via the BRS. 
So far, the technological options to achieve an intense seed THz light are severely limited. 
In addition, 
the current scheme enables to use inexpensive cathodes~\cite{cathod, cathod2, cathod3} and low-intensity infra-red lasers, instead of the accelerators and magnets~\cite{ freelaser, freelaser2, colson}. If successful, the cost of intense and powerful THz light source could be dramatically reduced while the intensity of the THz light is even higher than any current technology. The optimal regime of the current scheme is identified and estimated in this paper.

To begin with, consider two electron beams (denoted as 0 and 1) with electron density ($n_0 = n_1$) and  drift velocity ($v_{b0} > v_{b1}$). Considering a marginally relativistic electron velocity $v_{b0} $ and $v_{b1}$, the electron beam 1 has a drift velocity $ v_0 = (v_{b1} - v_{b0}) / (1- \beta_0 \beta_1) $ in the co-movnig frame with the electron beam, where $\beta_0 = v_{b0}/c$ and $\beta_1 = v_{b1}/c$. The electron beam density  in the co-moving frame is given as $n_0/\gamma_0 $ ($n_1/\gamma_0$), where $\gamma_0 =1/ \sqrt{1-\beta_0^2} $. Given an infra-red laser propagating in the same direction with the electron beams, the first objective is to excite the Langmuir waves via the two-stream instability, 
which  is appropriate for the BRS between the Langmuir wave and the infra-red laser. Denoting the frequency (wave vector) of the pump pulse in the laboratory as $\omega_{p0}$ ($k_{p0}$), the wave-vector of the infra-red laser in the co-moving frame is $ k_{p1} = k_{p0} \sqrt{(1-\beta_0)/(1+\beta_0)} $ becuase $\omega_{p0} \gg  \omega_{pe}$. As will be shown later, when  $c k_{p1} \gg 2 \omega_{pe} / \sqrt{\gamma_0} $, the wave vector of the appropriate Langmuir for the BRS in the co-moving frame
is  
\begin{equation}
  k_3 \cong 2 k_{p1} = 2  \left[\sqrt{\frac{(1-\beta_0)}{(1+\beta_0)}}\right]  k_{p0} \mathrm{.}
\label{eq:k_3}
\end{equation}
The Langmuir wave with the wave vector $k_{3} $ is hard to excite in an electron beam because of the high wave vector, $k_{p0} $. Therefore, it is optimal to utilize the excited Langmuir waves with the wave vector as high as possible. The estimation of the highest wave vector possible in exiciting the Langmuir wave is given below. The criteria of the two-stream instability would be that 1) there is a local maxima of the dielectric function $\epsilon$ as a function of $\omega$ for a fixed $k$ and 2) the value of the local maxima is less than zero. 

The longitudinal dielectric function of a classical plasma is   

\begin{equation} 
\epsilon(\mathbf{k}, \omega) = 1 + \frac{4 \pi e^2 }{k^2} \Sigma \chi_i \mathrm{.}
\end{equation} 
where the summation is over the group of particle species and  $  \chi_i $ is the particle susceptibility:

\begin{equation}
 \chi_i^C(k, \omega) = \frac{n_iZ_i^2}{m_i} \int \left[ \frac{ \mathbf{k} \cdot \mathbf{\nabla}_v f_i }{\omega - \mathbf{k} \cdot \mathbf{v} }\right]d^3 \mathbf{v} 
\end{equation} 
where $m_i$ ($Z_i$, $n_i$) is the particle mass (charge, density) and $f_i $ is the distribution with the normalization $\int f_i d^3 \mathbf{v} = 1$. For the case of our two group of electrons, it is given as 
\begin{equation} 
\epsilon =  1 + \left(\frac{4 \pi e^2}{k^2}\right) \left[ \chi_e^C(\omega, k) +  \chi_e^C(\omega-\mathbf{k}\cdot \mathbf{v}_0, k) \right] \mathrm{.}
 \label{eq:ele}
\end{equation} 
One example is shown in Fig~(\ref{fig1}) where a Langmuir wave is unstable to the two-stream instability.
 
\begin{figure}
\scalebox{0.3}{
\includegraphics[width=1.7\columnwidth, angle=270]{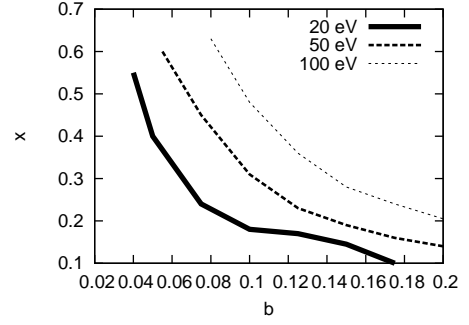}}
\caption{\label{fig2}
The threshold wave vector $k_C $ as a function of the drift velocity $v_0$. 
In this example, $n_0 = 10^{14} / \mathrm{cc} $ and  $\gamma_0 = 1.5$. 
The y-axis is  $ \mathrm{x} = k_C \lambda_{de} $ and the x-axis is $\mathrm{b} = v_0 / c$. 
Three cases when $T_e = 20 \  \mathrm{eV,} \ 50 \ \mathrm{eV,} \ 100 \ \mathrm{eV}$ are considered. 
There is no two-stream instability when 
 $v_0 / c < 0.04 $ for $T_e = 20  \ \mathrm{eV}$, 
 $v_0 / c < 0.055 $ for $T_e = 50 \  \mathrm{eV}$ and 
 $v_0 / c < 0.08 $ for $T_e = 100 \  \mathrm{eV}$
}
\end{figure}

\begin{figure}
\scalebox{0.3}{
\includegraphics[width=1.7\columnwidth, angle=270]{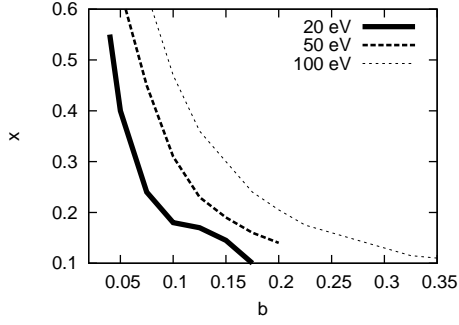}}
\caption{\label{fig3}
The threshold wave vector $k_C $ as a function of the drift velocity $v_0$. In this example, $n_0 = 10^{15} / \mathrm{cc} $ and  $\gamma_0 = 1.5$. The y-axis is  $ \mathrm{x} = k_C \lambda_{de} $ and the x-axis is $\mathrm{b} = v_0 / c$. Three cases when $T_e = 20 \  \mathrm{eV,} \ 50 \ \mathrm{eV,} \ 100 \ \mathrm{eV}$ are considered. There is no two-stream instability when $v_0 / c < 0.04 $ for $T_e = 20  \ \mathrm{eV}$, $v_0 / c < 0.06 $ for $T_e = 50 \  \mathrm{eV}$ and $v_0 / c < 0.08 $ for $T_e = 100 \  \mathrm{eV}$
}
\end{figure}

\begin{figure}
\scalebox{0.3}{
\includegraphics[width=1.7\columnwidth, angle=270]{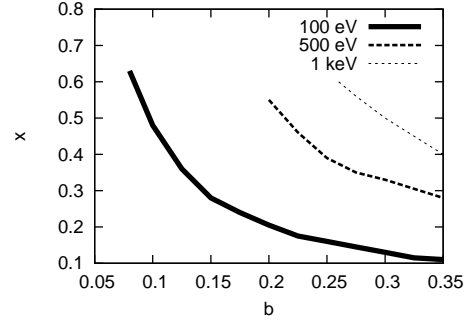}}
\caption{\label{fig4}
The threshold wave vector $k_C $ as a function of the drift velocity $v_0$. In this example, $n_0 = 10^{16} / \mathrm{cc} $ and  $\gamma_0 = 1.5$. The y-axis is  $ \mathrm{x} =  k_C \lambda_{de} $ and the y-axis is $\mathrm{b} = v_0 / c$. Three cases when $T_e = 100 \  \mathrm{eV,} \ 500 \ \mathrm{eV,} \ 1 \ \mathrm{keV}$ are considered. There is no two-stream instability when $v_0 / c < 0.08 $ for $T_e = 100  \ \mathrm{eV}$, $v_0 / c < 0.2 $  for $T_e = 500 \  \mathrm{eV}$ and $v_0 / c < 0.26 $ for $T_e = 1 \  \mathrm{keV}$
}
\end{figure}

The stability analysis is performed for the various beam density and temperature and illustrated in Figs.~(\ref{fig2}), (\ref{fig3}) and (\ref{fig4}). The analysis suggests that the optimal drift velocity $v_0 $ should be $2 < \sqrt{\gamma_0 }k v_0/\omega_{pe} < 4$ and the threshold condition for the two-stream instability is

\begin{equation} 
 k_3 \lambda_{de} =  2  \left[\sqrt{\frac{(1-\beta_0)}{(1+\beta_0)}}\right] \lambda_{de} < 0.5 \mathrm{,} \label{eq:two}
\end{equation}
where  $\lambda_{de} = \sqrt{\gamma_0 T_e/4 \pi n_0 e^2 } $. It can be deduced from Eq.~(\ref{eq:two}) that the lower the electron temperature, the higher the Langmuir wave vector can be. However, the electron temperature in the co-moving frame has the lower limit imposed by  the energy spread of the beam. If the beams has a energy spread $\delta E/E$, the electron temperature in the co-moving frame is comparable to $T_e \cong (\delta E/E)^2 m_e c^2$ from the fact that  $\delta v/c \cong \delta E / E$, where $\delta v $ is the velocity spread of the beam in the co-moving frame. If the energy spread is given as $ 0.01 < \delta E /E < 0.1$, the electron temperature in the co-moving frame would be $ 25 \ \mathrm{eV} < T_e < 2.5 \ \mathrm{keV} $.

The analysis of the BRS between the infra-red laser and the excited Langmuir wave is in order. The 1-D BRS three-wave interaction in the \textit{co-traveling} frame is~\cite{McKinstrie}:
\begin{eqnarray}
\left( \frac{\partial }{\partial t} + v_p \frac{\partial}{\partial x} + \nu_1\right)A_p  = -ic_p A_s A_3  \nonumber \mathrm{,} \nonumber \\ 
\left( \frac{\partial }{\partial t} + v_s \frac{\partial}{\partial x} + \nu_2\right)A_s  = -ic_s A_p A^*_3   \label{eq:2} \mathrm{,} 
\label{eq:2}
\end{eqnarray}
where $A_i= eE_{i1}/m_e\omega_{i1}c$  is the ratio of  the electron quiver velocity of the pump pulse ($i=p$) and the seed pulse ($i=s$) relative to the speed of light $c$, $E_{i1}$ is the electric field of the seed (pump) pulse, $A_3 = \delta n_1/n_1$ is the Langmuir wave amplitude, $ c_p = \omega_3^2/ 2 \omega_{p1}$ and $\omega_{p1}$ ($\omega_{s1}$) is the frequency of the pump (THz) pulse in the co-moving frame, and $\omega_3 \cong \omega_{pe} / \sqrt{\gamma_0} $ is the plasmon  frequency. In the co-moving frame, the infra-red laser satisfies the usual dispersion relation, $\omega_1^2 =  2  \omega_{pe}^2/\gamma_0 + c^2 k_1^2$, where $\omega_1$ and $k_1$ are the photon wave frequency and the corresponding vector. Denoting the wave vector and the frequency of the pump laser (seed pulse or THz light) in the co-traveling frame as  $k_{p1}$ and  $\omega_{p1} $ ($k_{s1} $ and $\omega_{s1} $) and their laboratory frame counterparts as $k_{p0}$ and $\omega_{p0} $ ($k_{s0} $ and  $\omega_{s0} $), the Lorentz transformation leads to the following relationship:
\begin{eqnarray} 
\omega_{p0} &=& \gamma_0 \left[ \sqrt{2\omega_{pe}^2/\gamma_0 + c^2 k_{p1}^2 } + vk_{p1} \right] \mathrm{,}  \label{eq:lorentz1} \\  \nonumber \\
k_{p0} &=&  \gamma_0 \left[ k_{p1} + \frac{\omega_{p1} }{c}  \frac{v_0}{c} \right] \mathrm{,} \label{eq:lorentz2} \\ \nonumber \\ 
 \omega_{s0} &=& \gamma_0 \left[ \sqrt{2\omega_{pe}^2/\gamma_0 + c^2 k_{s1}^2 } - 
vk_{s1} \right] \mathrm{,}  \label{eq:lorentz3} \\  \nonumber \\ 
k_{s0} &=&  \gamma_0 \left[ k_{s1} - \frac{\omega_{s1} }{c}  \frac{v_0}{c} \right]\mathrm{.} \label{eq:lorentz4} \\ \nonumber 
\end{eqnarray}
The energy and momentum conservation of Eq.~(\ref{eq:2}) leads to
\begin{eqnarray} 
 \omega_{p1} &=& \omega_{s1} + \omega_{3} 
\mathrm{,} \nonumber \\ 
 k_{p1} &=&  k_{s1} +k_3 \mathrm{,} \label{eq:cons} 
\end{eqnarray}
where $k_3$ is the wave vector of the plasmon. With a given pump frequency $\omega_{p0} $, $k_{p1} $ ($\omega_{p1} $) is determined from  Eq.~(\ref{eq:lorentz1}), $k_{s1} $ ($\omega_{s1}$) is determined from  Eq.~(\ref{eq:cons}), and $k_{s0} $ ($\omega_{s0}$) is determined from  Eqs.~(\ref{eq:lorentz3}) and (\ref{eq:lorentz4}). In the limiting case $ck_{s1} \gg \omega_3 $, $\omega_{s0} \cong  \sqrt{(1-\beta_0)/(1+\beta_0)} (\omega_{p1} - \omega_3)$ or   
\begin{equation} 
\omega_{s0} \cong  
\left[\frac{(1-\beta_0)}{(1+\beta_0)}\right] \left[ \omega_{p0} - 2 \omega_{pe} \sqrt{\gamma_0}\right] \mathrm{,}\label{eq:down}
\end{equation} 
where $\omega_{p1} \cong  \sqrt{(1-\beta_0)/(1+\beta_0)}  \omega_{p0}$ and $\omega_3 \cong \omega_{pe} /\sqrt{\gamma_0} $. Eq.~(\ref{eq:down}) describes the frequency down-shift of the visible-light pump  laser into the THz light, via the relativistic Doppler effect. For instance, if $\gamma_0= 1.5$, the down-shifted frequency would be $4.37$ THz for the CO2 laser whose frequency is 30 THz. From Eq.~(\ref{eq:lorentz4}) and Eq.~(\ref{eq:cons}), Eq.~(\ref{eq:k_3}) can be derived. 

It can be deduced from Eq.~(\ref{eq:2}) that the considerable part of the pump energy will be transferred to the seed pulse when $\sqrt{c_p c_s}  A_3 \delta t_b \cong 1$ or the mean-free-path of the BRS is
\begin{equation} 
l_b = \delta t c  \cong c (2  \sqrt{\omega_{s1}\omega_{p1}}   /\omega_3^2 ) (1/A_3) 
\mathrm{.} \label{eq:mean}
\end{equation}
On the other hand, the mean-free-path from  the Thomson scattering is  $ l_t \cong 1/n\sigma_t $ with $\sigma_t = (mc^2 / e^2)^2 $. When $n_1 \cong 10^{16} / \mathrm{cc} $, we estimate $l_t \cong 10^9 \ \mathrm{cm} $ and $l_b \cong (10^{-2} / A_3) (k_3c\omega_{pe}) /  \ \mathrm{cm} $. Even for $A_3 \cong 0.001 $, the THz light by the BRS is considerably stronger than the Thomson scattering or $l_t \gg l_b $.

The maximum conversion efficiency from the pump energy to the THz energy can be estimated as follows. Denote the total energy of the BRS pump laser (the seed laser) \textit{in the laboratory frame} as $\mathrm{E}_{p0} $ ($\mathrm{E}_{s0}$). \textit{In the co-moving frame}, the BRS pump energy is seen to be $\mathrm{E}_{p1}  \cong    \sqrt{(1-\beta_0)/(1+\beta_0)} \mathrm{E}_{p0}$. Considering the conversion efficiency in this co-moving as $\epsilon_1 $, the energy of the seed pulse is given as $\mathrm{E}_{s1} =  \epsilon_1  \sqrt{(1-\beta_0)/(1+\beta_0)} \mathrm{E}_{p1}$. This energy of the seed pulse is seen \textit{in the laboratory} to be $\mathrm{E}_{s0} = \epsilon_1   \sqrt{(1-\beta_0)/(1+\beta_0)} \mathrm{E}_{s1}$. Then, the conversion efficiency in the laboratory frame is

\begin{equation} 
\epsilon_0 =   \left[\frac{(1-\beta_0)}{(1+\beta_0)} \right] \epsilon_1  
\label{eq:conv} 
\end{equation}
The estimation of  $\epsilon_1$ in the co-moving frame is in order. If the plasmons are isotropically distributed, so does the radiation. However, only the photon in the direct opposite direction to the beam direction would be down-shifted to the THz range; The angular width, that are relevant to the THz, is $d \theta \cong  \sqrt{(1-\beta_0)/(1+\beta_0)}$ and the conversion efficiency would be   $\epsilon_1  \cong (d\theta)^2 \cong  (1-\beta_0)/(1+\beta_0) $. On the other hand, if the angular distribution of the plasmons are sharply peaked at $\theta = 0 $, the most of the pump pulse will be radiated into the opposite direction to the beam, in which case  $\epsilon_1 \cong 1 $. From the above consideration, the optimal conversion efficiency is estimated as  

\begin{equation}
\left[\frac{(1-\beta_0)}{(1+\beta_0)} \right]^2   <  \epsilon_0  <  
 \left[\frac{(1-\beta_0)}{(1+\beta_0)} \right] \mathrm{.}  \label{eq:eps}
\end{equation}

A few examples are provided below. As the first example, consider the case when $n_0 = 10^{14} \ /\mathrm{cc} $ and $\gamma_0 = 2.0$. 
When $T_e = 20 \ \mathrm{eV}$,
 $k_3 \lambda_{de} \cong 0.5 $ ($k_3 \lambda_{de} \cong 1.0 $) 
if the wave length of the infra-red laser is  $\lambda = 20 \ \mu \mathrm{m}$
 ($\lambda = 10 \ \mu \mathrm{m}$). 
Only the infra-red laser with $\lambda = 20 \ \mu \mathrm{m}$ 
would be suitable for the THz generation  since the Langmuir wave for the BRS is unstable only 
for the laser with  $\lambda = 20 \ \mu \mathrm{m}$ and the emitted light has the frequency of 2.15 THz.
If $T_e = 50 \ \mathrm{eV} $,  $k_3 \lambda_{de} \cong 0.98 $ ($k_3 \lambda_{de} \cong 1.96 $) if  the wave length of the infra red laser as $\lambda = 20 \ \mu \mathrm{m}$
 ($\lambda = 10 \ \mu \mathrm{m}$). The Langmuir wave for the BRS is stable 
to the two-stream instability and both lasers are unsuitable for the THz generation. 

As the second example, 
 consider the case when $n_0 = 10^{15} \ /\mathrm{cc} $ and $\gamma_0 = 1.5$. 
When $T_e = 20 \ \mathrm{eV}$,
 $k_3 \lambda_{de} \cong 0.19 $ ($k_3 \lambda_{de} \cong 0.39 $) 
if  the wave length of the infra-red laser is $\lambda = 20 \ \mu \mathrm{m}$
 ($\lambda = 10 \  \mu \mathrm{m}$). 
The Langmuir wave for the BRS is unstable when $0.06 < v_0/c < 0.12 $ ($0.06 < v_0/c < 0.11 $). Both lasers are suitable for the THz generation and 
the emitted light has the frequency of 4.37 THz (2.18 THz) for 
 $\lambda = 10 \ \mu \mathrm{m}$ ($\lambda = 20 \ \mu \mathrm{m}$).
When $T_e = 50 \ \mathrm{eV}$,
   $k_3 \lambda_{de} \cong 0.31 $ ($k_3 \lambda_{de} \cong 0.62 $) 
if  the wave length of the infra-red laser is $\lambda = 20 \ \mu \mathrm{m}$
 ($\lambda = 10 \  \mu \mathrm{m}$). 
The Langmuir wave for the BRS is unstable (stable) when $0.06 < v_0/c < 0.12 $ for $\lambda = 20 \ \mu \mathrm{m}$ ($\lambda = 10 \ \mu \mathrm{m}$). 
The laser with $\lambda = 20 \ \mu \mathrm{m}$ is suitable for the THz generation and 
the emitted light has the frequency of 2.18 THz.
When $T_e = 100 \ \mathrm{eV}$,
   $k_3 \lambda_{de} \cong 0.43 $ ($k_3 \lambda_{de} \cong 0.86 $) 
if  the wave length of the infra-red laser as $\lambda = 20  \mu \mathrm{m}$
 ($\lambda = 10  \ \mu \mathrm{m}$). 
The Langmuir wave for the BRS is unstable (stable) when $0.08 < v_0/c < 0.11 $ for $\lambda = 20 \  \mu \mathrm{m}$ ($\lambda = 10\  \mu \mathrm{m}$). 
The laser with $\lambda = 20 \ \mu \mathrm{m}$ is suitable for the THz generation and the emitted light has the frequency of 2.18 THz. 

As the third example, 
 consider the case when $n_0 = 10^{16} \ /\mathrm{cc} $ and $\gamma_0 = 1.5$. 
When $T_e = 100 \ \mathrm{eV}$,
 $k_3 \lambda_{de} \cong 0.13 $ ($k_3 \lambda_{de} \cong 0.27 $) 
if  the wave length of the infra-red laser is $\lambda = 20 \ \mu \mathrm{m}$
 ($\lambda = 10 \ \mu \mathrm{m}$). 
The Langmuir wave for the BRS is unstable when $0.06 < v_0/c < 0.3 $ ($0.06 < v_0/c < 0.15 $). Both lasers are suitable for the THz light generation and the emitted light has the frequency of 4.37 THz (2.18 THz) for 
 $\lambda = 20 \ \mu \mathrm{m}$ ($\lambda = 10 \ \mu \mathrm{m}$). 
When $T_e = 500 \ \mathrm{eV}$,
   $k_3 \lambda_{de} \cong 0.31 $ ($k_3 \lambda_{de} \cong 0.62 $) 
if the wave length of the infra-red laser is $\lambda = 20 \ \mu \mathrm{m}$
 ($\lambda = 10 \  \mu \mathrm{m}$). 
The Langmuir wave for the BRS is unstable (stable) when $0.2 < v_0/c < 0.32 $ for $\lambda = 20 \ \mu \mathrm{m}$ ($\lambda = 10 \ \mu \mathrm{m}$). 
Both lasers are suitable for the THz light generation and 
 the emitted light has the frequency of 4.37 THz (2.18 THz) for 
 $\lambda = 20 \ \mu \mathrm{m}$ ($\lambda = 10 \ \mu \mathrm{m}$). 
When $T_e = 1000 \ \mathrm{eV}$,
   $k_3 \lambda_{de} \cong 0.43 $ ($k_3 \lambda_{de} \cong 0.86 $) 
if  the wave length of the infra-red laser is $\lambda = 20 \ \mu \mathrm{m}$
 ($\lambda = 10 \ \mu \mathrm{m}$). 
The Langmuir wave for the BRS is unstable (stable) when $0.26 < v_0/c < 0.34 $ for $\lambda = 20 \ \mu \mathrm{m}$ ($\lambda = 10 \  \mu \mathrm{m}$). 
The laser with $\lambda = 20 \ \mu \mathrm{m}$  is suitable for the THz light generation and  the emitted light has the frequency of 2.18 THz.

In summary, we propose a new scheme of THz light source based on the two-stream instability and the backward Raman scattering. The excitation of the Langmuir wave via the two-stream instability and the interaction of the co-propagating infra-red laser with the excited Langmuir wave results in the THz light emitted in the opposite direction to the electron beam.
The current scheme proposes a plausible solution for generating a strong seed pulse. 
Furthermore, the excited Langmuir waves can be  used directly for the BRS, bypassing the seed amplification via the ponderomotive interaction between the seed pulse and the laser, resulting that 
 the intensity requirement of a visible light laser (especially the infra-red laser) is lowered considerably. 
The current invention  shifts the paradigm of the BRS utilization  from the well-tested compression technology~\cite{Fisch} to the THz light generation technology, and   
 the proposed frequency down-shift, in conjunction with the Langmuir wave instability and the BRS,  
creates an attractive opportunity 
   for an ultra-intense THz light source.  
The estimation of the conversion efficiency and the constraint of the scheme are analyzed in this paper. 

The discussion on the available electron beams follows. The possible electron temperature of the electron beams in the co-moving frame would be $25 \ \mathrm{eV} < T_e < 2.5 \ \mathrm{keV} $. In order for the new scheme to be plausible, the electron temperature should be $T_e < 50 \ \mathrm{eV}$ ($T_e < 500  \ \mathrm{eV}$, $T_e < 2  \ \mathrm{keV}$) when $n_0 = 10^{14} \  /\mathrm{cc} $ ($n_0 = 10^{15} \  /\mathrm{cc} $, $n_0 = 10^{16} \  /\mathrm{cc} $). The constraint on the electron temperature will be more more relaxed if the density is higher; $T_e <5 \ \mathrm{keV}$ when $n_e = 10^{17} \  /\mathrm{cc} $. With the current cathode technologies~\cite{cathod, cathod2, cathod3}, the electron density up to $10^{14} - 10^{15} \  /\mathrm{cc} $ is achievable with the beam energy of a few hundred keV and can be increased from some focusing technologies. While it is desirable to  use electron beams from cathodes, it is also possible to obtain the electron beams from the laser-plasma interaction. With the current technologies~\cite{monoelectron, ebeam,lbeam} developed for the inertial confinement fusion~\cite{tabak,sonprl,sonpla, sonpla2, sonpla3, sonchain}, the electron density could be as high as $10^{23} \ / \mathrm{cc} $ with $\gamma_0 \cong 100 $, but the beams required in our scheme has much lower energy (less dense) so that the cost could be cheaper. 

There have been researches on the soft x-ray or THz light source based on the direct BRS~\cite{sonbrs, brs, brs2, brs3, drake}. In author's  previous research~\cite{sonbrs}, the direct use of the BRS and the Doppler's effect has been proposed. There are mainly two differences between the current scheme and the previous one. First, the Langmuir wave is excited by the two-stream instability rather than by an intense laser. Second, the infra-red laser are preferred to the visible light in the current scheme. The use of  the infra-red laser would not be plausible in the previous scheme because the intensity of the infra-red laser required by the previous scheme can not be achieved under the current laser technologies. Due to the low frequency of the infra-red laser, the required kinetic energy of the electron beam is low, resulting that the electron beam can be generated by a cheaper cathode or conventional (laser-based) accelerator.

In this paper, it is assumed that the Langmuir wave for the BRS will be excited as long as it it unstable to the two-stream instability. However, the future work needs to check through the simulation and the theoretical analysis how intense the excited Langmuir wave is. The full adequacy of the current scheme can be further validated by the study along this line, which is beyond the scope of this paper. The plausibility study of the current scheme in a fully relativistic electron beam and the ND:YAG laser might be another interesting research topic.
\bibliography{tera2}

\end{document}